\documentclass[floats,twocolumn,aps,prb,longbibliography]{revtex4-2}

\usepackage{graphicx}
\usepackage{amsfonts}
\usepackage{amssymb}
\usepackage{amsbsy} 
\usepackage{amsmath}\usepackage[colorlinks = true,
            linkcolor = magenta,
            urlcolor  = magenta,
            citecolor = magenta,
            anchorcolor = magenta]{hyperref}

\makeatletter
\newcommand*{\balancecolsandclearpage}{%
  \close@column@grid
  \clearpage
  \twocolumngrid
}
\makeatother

\def\up{{u_+}}
\def\um{{u_-}}

\def\E{\mathcal E}

\def\K{\mathcal K}

\def\P{\mathcal P}

\begin{document}

\title{Logarithmic divergent specific heat from high-temperature series expansions: application to the two-dimensional XXZ Heisenberg model}

\author{M. G. Gonzalez$^1$, B. Bernu$^1$, L. Pierre$^2$, and L. Messio$^{1,3}$}

\affiliation{$^1$Sorbonne Universit\'e, CNRS, Laboratoire de Physique Th\'eorique de la Mati\`ere Condens\'ee, LPTMC, F-75005 Paris, France}
\affiliation{$^2$Paris X, F-92000 Nanterre, France}
\affiliation{$^3$Institut Universitaire de France (IUF), F-75005 Paris, France}

\begin{abstract} 
We present an interpolation method for the specific heat $c_v(T)$, when there is a phase transition with a logarithmic singularity in $c_v$ at a critical temperature $T=T_c$. The method uses the fact that $c_v$ is constrained both by its high temperature series expansion, and just above $T_c$ by the type of singularity. 
We test our method on the ferro and antiferromagnetic Ising model on the two-dimensional square, triangular, honeycomb, and kagome lattices, where we find an excellent agreement with the exact solutions. We then explore the XXZ Heisenberg model, for which no exact results are available.
\end{abstract}

\maketitle 

\section{Introduction}

The study of finite-temperature phase transitions is of great importance in condensed matter physics\cite{Li21}, and in particular in the field of quantum magnetism. When lowering the temperature from a paramagnetic state, a transition may occur at a critical point where an order parameter arises due to the breaking of a given symmetry of the Hamiltonian. The interest in studying these transitions is twofold: on the theoretical side, the classification of the different phase transitions according to their universality classes and the understanding of the parameters that play a key role in the change from one phase to the next. On the experimental side, reliable methods are needed to contrast measurements with theoretical predictions. And even though the Mermin-Wagner theorem states that the breaking of continuous symmetries at finite temperature can only occur in three-dimensional systems\cite{Mermin66}, two-dimensional systems can still have finite-temperature phase transitions through the breaking of a discrete symmetry, as in the case of the Ising model.

The Ising model was first proposed in 1920 by W. Lenz in an attempt to explain ferromagnetism\cite{Lenz20}, and his student E. Ising found that the exact solution in one dimension has no phase transition whatsoever\cite{Ising25}. Many years later it was proven that a paramagnetic to ferromagnetic phase transition does indeed occur at finite temperature in the two-dimensional model on the square lattice\cite{Peierls36, Kramers41}, finally exactly solved by L. Onsager in 1944\cite{Onsager44}. Using different transformations this result was extended to several two-dimensional lattices\cite{Wannier50, Utiyama51, Kano53, Naya54, Fisher59, Plechko98, Urumov02, Strecka06}, such as the triangular lattice\cite{Wannier50}, the kagome lattice\cite{Kano53}, and the honeycomb lattice\cite{Naya54}. A detailed summary of the critical temperatures on Archimedean and Laves lattices is given in Ref. \cite{Codello10}. 

But the Ising model is not only important in the understanding of ferromagnetism. In fact, the critical exponents of the thermodynamic functions near the critical point form the Ising universality class, and there are several systems (not only ferromagnetic) which behave similarly in the presence of a phase transition\cite{Messio08, Domenge08}. In the field of quantum magnetism, the Mermin-Wagner theorem precludes the existence of finite-temperature phase transitions in two-dimensional systems when only continuous symmetries are broken in the ground state, like in most Heisenberg magnets; but not when the ground state has a discrete degeneracy\cite{Villain77}. An important example of this is the case of the Heisenberg model with nearest and next-nearest neighbor interactions in the square lattice\cite{Chandra90, Weber03, Singh03, Capriotti04}.  Chandra, Coleman and Larkin proposed that in this model a transition could occur due to the emergence of an effective Ising order parameter\cite{Chandra90}. This transition was confirmed several years later for the classical model on the square lattice by Monte Carlo simulations\cite{Weber03}, showing that the critical exponents were those of the Ising model. And although there was some doubt about the extension to the $S=1/2$ quantum case, it was realized shortly after that the Ising transition occurs for any value of $S$\cite{Singh03, Capriotti04}. 

The theoretical calculation of critical properties is then of central importance to understand the relevant order parameters involved in phase transitions. When no exact result is available, numerical methods have to be implemented. Classical and quantum Monte Carlo (QMC) methods work naturally at finite temperatures and grant access to the calculation of the critical exponents and temperatures through finite-size scaling\cite{Weber03}, but fail in the quantum case when there is frustration due to the sign problem. For quantum systems, exact diagonalization\cite{Lauchli19} and tensor methods\cite{Poilblanc21} are used on frustrated systems but present more difficulties when it comes to large lattice sizes and/or finite temperatures. 

Finally, high-temperature series expansion (HTSE) methods operate directly in the thermodynamic limit but cannot reach very low temperatures, a limitation which becomes more restrictive when there are singularities in the thermodynamic functions due to the presence of a finite-temperature phase transition. When there is no finite-temperature phase transition, methods have been proposed to extrapolate the HTSE of thermodynamic functions down to $T=0$\cite{Roger98, Bernu01, Schmidt17, Derzhko20}, relying on some knowledge about the ground state (such as the type of order and ground-state energy). These methods have proven to be useful for understanding experiments\cite{Misguich03, Misguich05, Faak12, Bernu15, Orain17, Bernu20}. But, apart from an exploratory extension of the entropy method presented in Ref. \cite{Grison20}, no other method has been developed to interpolate the thermodynamic functions between the high-temperature limit and the critical behavior at $T=T_c$.

In this paper, we present an interpolation method for the specific heat, $c_v$, based on the HTSE and the existence of a phase transition with a logarithmic singularity. We test it on several two-dimensional ferro and antiferromagnetic Ising models and then apply it to Heisenberg models with Ising anisotropy (XXZ models), all of which have a logarithmic divergence in $c_v$. The non-universal parameters characterizing the transition are obtained in a self-consistent way by looking for the greatest number of coinciding Pad\'e approximants (PA) of our proposed function. The reconstructed $c_v$, in the temperature range $[T_c,\infty]$, is in good quantitative agreement with the exact solution. For the XXZ model, we are able to evaluate $T_c$ over a wide range of anisotropies, finding a good agreement with the quantum Monte Carlo calculations available in the literature. 

The article is organized as follows. In Sec.~\ref{secMod}, we present the Hamiltonian for the XXZ model and some results derived from the exact solution in the Ising limit for the critical parameters of $c_v$. Sec.~\ref{secMet} is devoted to the interpolation method. In Sec.~\ref{secRes}, we present the results, first for the Ising limit and then for the XXZ model. Finally, the conclusions and perspectives are given in Sec.~\ref{secCon}.

\section{XXZ Model}
\label{secMod}

The $S=1/2$ XXZ Hamiltonian can be written as
\begin{equation}
	\mathcal{H} = J \sum_{\langle i j \rangle} \left[ {S}_i^z {S}_j^z + \Delta \ \mathbf{S}_i^\perp \cdot \mathbf{S}_j^\perp \right],
\label{eq01}
\end{equation}
where $\langle i j \rangle$ represents the nearest neighbors in a two-dimensional lattice, $S^z$ is the $z$-component of the spin and $\mathbf{S}^\perp = (S^x, S^y)$ represents the $x$ and $y$ components. 
In the following, we set $J = 1$ for the antiferromagnetic cases and $J=-1$ for the ferromagnetic ones, and $\Delta$ goes from 0 (Ising) to 1 (Heisenberg).

The ferromagnetic Ising model (and antiferromagnetic on bipartite lattices), $\Delta =0$, is known to exhibit a singularity in $c_v$ described above $T_c$ as
\begin{equation}
	c_v^s(\beta) = A\ln \left(1 - \frac{\beta}{\beta_c} \right)
\label{eqsing}
\end{equation}	
where $\beta=1/T$ is the inverse temperature ($\beta_c=1/T_c$). Throughout this article, we focus only on $T>T_c$ ($\beta<\beta_c$).

Onsager's exact solution of the Ising model on the square lattice has been extended to other lattices\cite{Plechko98}. The exact expressions of $c_v$ for the square, triangular, honeycomb and kagome lattices, and the corresponding singularity parameters are given in App.~\ref{app1}. We give in table \ref{tab0} the expressions of $\beta_c$, $A$ (see Eq.~\eqref{eqsing}), and the shift $B$ at $T_c$ defined by
\begin{equation}
	B=\lim_{\beta\to\beta_c} \left(c_v(\beta)-c_v^s(\beta)\right)
\label{eq-def-B}
\end{equation}	
\begin{table}[!t]
\centering
\renewcommand{\arraystretch}{2.3}
\begin{tabular}{l|c|c|r@{+}ll}
\hline
Lattice & $\beta_c$ & $-A$ & \multicolumn{2}{c}{$B/A$} \\
\hline
Square & $\ln(3\!+\!2\sqrt{2}) $  & $\displaystyle\frac{\beta_c^2}{2\pi} $ & $\displaystyle\frac{\pi}{4}$&$\displaystyle\ln \frac{e \beta_c}{4\sqrt{2}} $ \\
Triangular & $\ln 3$ & $\displaystyle\frac{3\sqrt{3}\beta_c^2}{4\pi}$  & $\displaystyle \frac{\pi}{2\sqrt{3}}$&$\displaystyle\ln \frac{e \beta_c}{4}$ \\
Honeycomb & $2\ln(2\!+\!\sqrt{3})$ & $\displaystyle\frac{\sqrt{3} \beta_c^2 }{8\pi}$ & $\displaystyle\frac{\pi\sqrt{3}}{9}$&$ \displaystyle\ln \frac{e\beta_c}{4\sqrt{3}}$ \\
Kagome & $\ln(3\!+\!2\sqrt{3})$ & $\displaystyle\frac{\sqrt{3}\beta_c^2}{4\pi}$ &  $\displaystyle \frac{\pi (\sqrt{3}\!-\!1)}{6}$&$\displaystyle \ln \frac{e\beta_c}{4} $
\end{tabular}
\caption{Singularity parameters $\beta_c$, $A$, and $B$ obtained from the exact solution of the ferromagnetic Ising model (see equations \eqref{eqsing}, \eqref{eq-def-B} and App.~\ref{app1}). $e$ stands for the Euler's constant.}
\label{tab0}
\end{table}

When the parameter $\Delta$ is turned on, the Ising transition survives and $c_v$ continues to have the same kind of divergence, while the critical temperature $T_c$ decreases. At the Heisenberg point, $\Delta =1$, there is no finite-temperature phase transition. Around the Heisenberg point ($\Delta\simeq 1$), the critical temperature behaves as $1/\ln|1-\Delta|$\cite{Hikami80, Kawabata86, Ding92}. This happens for both classical and quantum spins. For $\Delta <1$, which is the case we are focusing on, the phase transition is always of the Ising universality class\cite{Ding92, Loh85, Sariyer19}, whereas for $\Delta >1$ the transition is of the Kosterlitz-Thouless type\cite{Kosterlitz73, Hikami80,Kawabata82, Loh85,Ding92,  Lee05, Sariyer19}.

\section{Interpolation Method}
\label{secMet}

In this section we present a method to interpolate between the high-temperature limit and the critical temperature $T_c$. From the HTSE of the free energy, $f$, we evaluate the series expansions of the thermodynamic quantities around $\beta = 0$. The $f$-HTSE reads
\begin{equation}
	\beta f = -\ln 2 - \sum_{i=1}^n \frac{e_i}{2^i i!} \beta^i + O(\beta^{n+1})
\end{equation}
where $e_i$ are polynomials of $J$ and $\Delta$. In the Ising limit, where exact formulae are available, the coefficients can be calculated up to arbitrary orders, but when the exact partition function is unknown the HTSE is typically known up to orders 15 to 20 depending on lattice connectivity and the Hamiltonian. The convergence radius of the series is determined by the singularity closest to the origin in the complex plane of $\beta$. When the singularity appears on the real positive axis, a finite-temperature phase transition occurs. In practice, the series generally converge down to $T \approx |J|$ while Pad\'e approximants (PA) allow this range to be extended to about $0.5|J|$, in the absence of phase transitions. 

However, even if the raw HTSE or the PAs are unreliable close to a phase transition, the ratio or the Dlog Pad\'e methods allow to obtain accurate values of $\beta_c$\cite{Oitmaa96,Butera97,Muller15,Richter15,Oitmaa06,Lohmann14,Kuzmin19}. The ratio method is a direct estimate of the convergence radius of the series for a given thermodynamic function\cite{Oitmaa96, Oitmaa06,Lohmann14,Kuzmin19}. This method is usually used with the susceptibility $\chi(\beta)$, where the critical exponent $\gamma$ is greater than 1, but tends to fail for quantities such as the specific heat $c_v$ or the entropy $s$\cite{Oitmaa96} where the critical exponents are small or negative. The Dlog Pad\'e method\cite{Oitmaa06} evaluates the poles and residues of the PAs of a given function that possesses a simple pole in theory. This function is often the derivative of $\ln \chi(\beta)$, and this method gives very good results for $\beta_c$ and the critical exponent $\gamma$ as there are generally several PAs with the same pole and residue, up to a good precision\cite{Oitmaa96, Oitmaa06,Oitmaa18, Kuzmin20}. But both methods require some knowledge of the magnetic order, and the calculation of the corresponding susceptibility (either ferromagnetic, antiferromagnetic, or other).

The previous methods work because the series already hold information about the singularity at the lowest orders. Our goal is to use this information to develop a method that works for $c_v$ and provides reliable results over the whole range of temperatures down to $T_c$. We assume that $c_v$ around $T_c$ can be well described by the singular behavior from Eq.~\eqref{eqsing}, with a good choice of the parameters $A$ and $\beta_c$. Then we propose to subtract the singular behavior of $c_v$, and define a function $R(\beta)$
\begin{equation}
	R(\beta) = c_v(\beta) - c_v^s(\beta)
\label{eqreg}
\end{equation}
which behaves smoothly for $T\gtrsim T_c$. Note that the constant $B$ verifies $B=R(\beta_c)$ (see Eq.~\eqref{eq-def-B}). In practice we do not have access to the exact $c_v$ and we only have the $c_v$-HTSE to order $n$. The HTSE of the singular part $c_v^s(\beta)$ can be easily obtained up to order $n$. So, finally, we can obtain the $R$-HTSE to order $n$, which depends on parameters $\beta_c$ and $A$. This regular function $R$ is the one that we are going to use for the interpolation method. If $\beta_c$ or $A$ are wrongly chosen, the singularities in $c_v^s$ and $c_v$ will not cancel out and $R(\beta)$ will not behave regularly around $\beta_c$. 

Now we define the procedure to measure the quality of a given $R$-HTSE, defined by a set of parameters $\{\beta_c, A\}$. From the $R$-HTSE to order $n$ we get the standard $n+1$ PAs, defined with the convention that the denominator polynomial starts with 1. Then, we discard all the PAs with real poles in the range $[0,\beta_c]$ and we are left with $N_\P$ physical PAs, denoted by $\P_i$ with $i=1..N_\P$. The next step is to count how many of them coincide at $\beta_m = \delta \beta_c$ with $\delta > 1$, because a regular function at $\beta_c$ should be regular beyond the critical point. Throughout the article, we use $\delta = 1.05$. Since the $R$-HTSE starts at order 1 in $\beta$ (the $c_v$-HTSE starts at order 2, and the $c_v^s$-HTSE starts at order 1), we discard the PA with a constant numerator (order 0). We then define a quality function $Q$ as
\begin{equation}
	Q^2 = \frac{2}{(n-1)n} \sum_{i=1}^{N_\P} \sum_{j=1}^{i-1} M_\epsilon\left(\frac{\P_i(\beta_m)-\P_j(\beta_m)}{\overline{F}(\beta_c)}\right)
\label{eq-def-Q}
\end{equation}
where $M_\epsilon(x) = 1/(1+(x/\epsilon)^8)$ and $\overline{F}(\beta_c)=\frac12(\P_i(\beta_c)+\P_j(\beta_c))$. The function $M_\epsilon(x)$ is chosen to be near zero when $x>\epsilon$ and close to 1 otherwise, and we choose $\epsilon = 0.005$. Then, $Q$ represents the proportion of coinciding PAs. As we will show below, $Q$ takes large values only in small regions of the parameter space and we will consider peaks of $Q(A,\beta_c)$ greater than 0.5 as reliable results (a majority of coinciding PAs). The best set of parameters $\{\beta_c,A\}$ is defined by the maximum in $Q(A,\beta_c)$. We define the uncertainties on this best set of parameters as the width of the peaks.

Once the best set of parameters $\{\beta_c,A\}$ is found, we can reconstruct the specific heat $c_v^r$ by adding the singularity to the regular part. For that, we replace $R(\beta)$ by one of its coinciding PAs (one contributing to $Q$): $\P_i$; and we put explicitly the singular part in Eq. \ref{eqreg}:
\begin{equation}
c_v^r(\beta) = \P_i(\beta) + A \ln \left(1 - \frac{\beta}{\beta_c} \right)
\label{eqrecon}
\end{equation}
In the limit $n\to \infty$, the reconstructed $c_v^r$ should converge to the exact one.

\section{Results}
\label{secRes}

\begin{figure}[!t]
    \begin{center}
        \includegraphics*[width=0.47\textwidth]{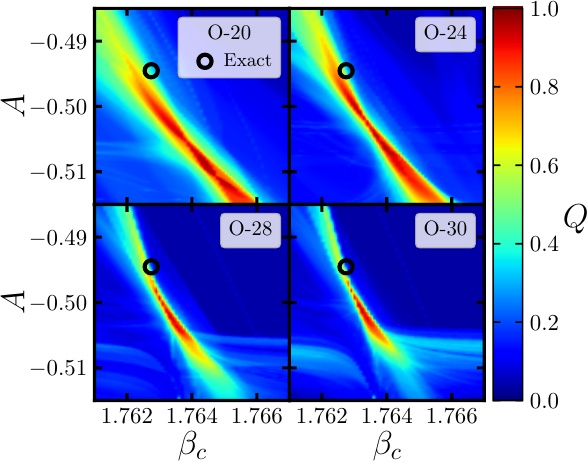}
        \caption{The quality $Q$ (see Eq.~\eqref{eq-def-Q}) versus $\beta_c$ and $A$ (see Eqs.~\eqref{eqsing} and \eqref{eq-def-B}) for the square lattice ferromagnetic Ising model using the $c_v$-HTSE at orders 20 (top-left), 24 (top-right), 28 (bottom-left), and 30 (bottom-right). The black circles indicate the Onsager's exact solution (see Table \ref{tab0}). Note that $Q$ reaches values close to 1.}
        \label{fig1}
    \end{center}
\end{figure}

In this section we present the results obtained with the method presented in the previous section for different lattices and models. Whenever possible, we compare our results with exact results or calculations by other authors using alternative methods. 

\subsection{Ferromagnetic Ising model on the square lattice}

We compute the quality $Q$ (see Eq.~\eqref{eq-def-Q}) for several orders of the HTSE on a grid of values in the plane $\beta_c$-$A$. We are scanning in steps of $10^{-4}$ both in $\beta_c$ and $A$ in a wide range around the exact values. It is important to keep small steps, because the domain where the quality $Q$ reaches its maximum is generally narrow as it can be seen in figure \ref{fig1}, where the black circles represent the exact values. The best sets of parameters have quality values around 0.95, which means that almost all of the PAs coincide. In addition, we find that the quality drops rapidly away from the best parameters.

\begin{figure}[!t]
\begin{center}
\includegraphics*[width=0.45\textwidth]{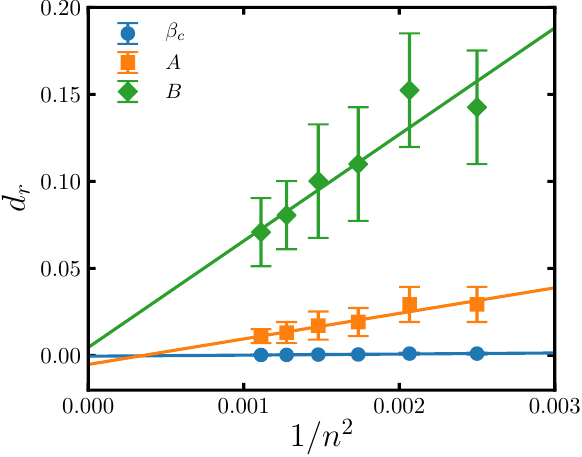}
\caption{Relative differences of the values of $\beta_c$, $A$ and $B$ extracted from Table~\ref{tab1} compared to to the exact ones as a function of $1/n^2$, where $n$ is the HTSE order. Lines are linear fits through the points taking into account the uncertainties. $d_r = (V-V^\text{exact})/V^\text{exact}$, where $V=\beta_c$, $A$, and $B$.}
\label{fig2}
\end{center}
\end{figure}

As the order of the HTSE increases, we see that the best-quality domain is narrower and gets closer to the exact point, as shown in table \ref{tab1}. In figure \ref{fig2}, we see that the parameters go to roughly the exact values like $1/n^2$, where $n$ is the HTSE order. Note that already at order 20 the results are close to the exact ones (see table \ref{tab1}) . 

\begin{table}[!t]
\centering
\begin{tabular}{c|l|l|l}
\hline
Order & \multicolumn{1}{c|}{$\beta_c$} & \multicolumn{1}{c|}{$A$} & \multicolumn{1}{c}{$B$} \\
\hline
20 & 1.7645(10) & $-0.509(5)$ & $-0.350(10)$ \\
22 & 1.7645(10) & $-0.509(5)$ & $-0.353(10)$\\
24 & 1.7636(6) & $-0.504(4)$ & $-0.340(10)$ \\
26 & 1.7635(6) & $-0.503(4)$ & $-0.337(10)$ \\
28 & 1.7632(3) & $-0.501(3)$ & $-0.331(6)$ \\
30 & 1.7631(3) & $-0.500(2)$ & $-0.328(6)$ \\
\hline
Exact & 1.762747... & $-0.49453$... &  $-0.30631$...
\end{tabular}
\caption{Results for the parameters $\beta_c$, $A$, and $B$ versus the HTSE order for the square lattice ferromagnetic Ising model, compared to the exact solution (see Table~\ref{tab0}).}
\label{tab1}
\end{table}

\begin{figure}[!t]
\begin{center}
\includegraphics*[width=0.45\textwidth]{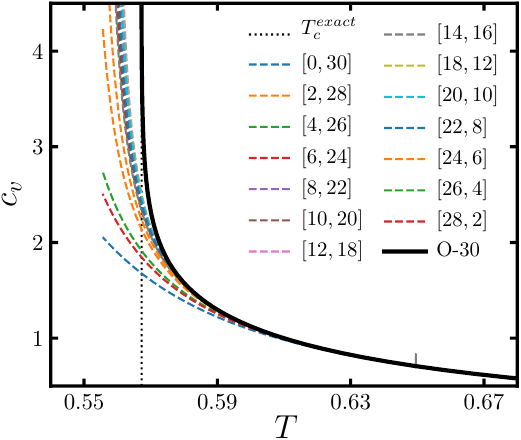}
\caption{The specific heat $c_v$ as a function of the temperature $T$ for the square lattice ferromagnetic Ising model: full line stands for the reconstructed $c_v^r$ from our method. The dashed lines are the the Pad\'e approximants of the raw $c_v$. The vertical dotted line indicates the exact value of $T_c$.}
\label{fig3}
\end{center}
\end{figure}

Using the best parameters $\{\beta_c,A\}$ presented in table \ref{tab1}, we reconstruct the specific heat $c_v^r$ using Eq. \ref{eqrecon}. Figure \ref{fig3} shows $c_v^r(T)$ built from the HTSE at order 30, where only a single PA does not coincide. For comparison the dashed lines represent the PAs of the raw HTSE of $c_v$. The present method shows a clear improvement over the raw HTSE or its PAs.

\begin{figure}[!t]
\begin{center}
\includegraphics*[width=0.45\textwidth]{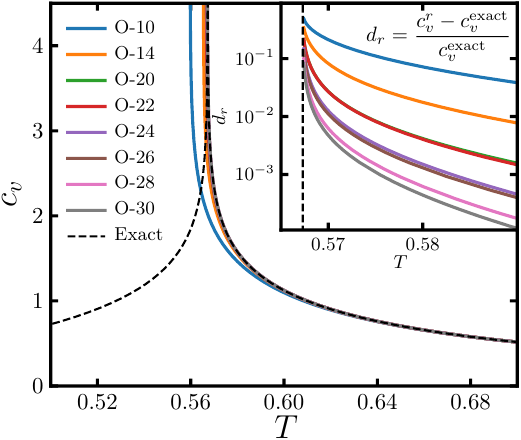}
\caption{The reconstructed $c_v^r$ versus the temperature $T$ for the square lattice ferromagnetic Ising model using $c_v$-HTSE orders from 10 to 30. The black dashed line indicates the exact solution. The inset shows the relative difference between our results and the exact result.}
\label{fig4}
\end{center}
\end{figure}

In figure \ref{fig4} we show the results for the reconstructed $c_v^r$ obtained from the HTSE at orders from 10 to 30 (solid lines), compared with the exact result (dashed line). Even though $c_v^r$ at order 10 is visibly different from the exact one, the results improve quickly as the order increases. The inset of figure \ref{fig4} shows that the relative differences with respect to the exact result decrease significantly as the order increases. Since the  $\beta_c$ evaluated at each order is slightly greater than the exact value, the relative differences tend to 1 around the exact critical temperature. We see that a good agreement with the exact results is already obtained at order 14, even if the values of $A$ and $B$ may differ from the exact ones. This means that the precision in the parameters $A$ and $B$ is not as important as the precision in $\beta_c$ in order to have a good representation of $c_v$.

\subsection{Ferromagnetic Ising model on other lattices}

\begin{table*}[t]
\centering
\begin{tabular}{c|l|l|l||l|l|l||l|l|l}
\hline
& \multicolumn{3}{c||}{Triangular} & \multicolumn{3}{|c||}{Honeycomb} & \multicolumn{3}{|c}{Kagome} \\
\hline
\multicolumn{1}{c|}{Order} & \multicolumn{1}{c|}{$\beta_c$} & \multicolumn{1}{c|}{$A$} & \multicolumn{1}{c||}{$B$} & \multicolumn{1}{c|}{$\beta_c$} & \multicolumn{1}{c|}{$A$} & \multicolumn{1}{c||}{$B$} & \multicolumn{1}{c|}{$\beta_c$} & \multicolumn{1}{c|}{$A$} & \multicolumn{1}{c}{$B$} \\
\hline
20 & 1.0998(4) & $-0.521(4)$ & $-0.385(10)$ & 2.6395(20) & $-0.484(5)$ & $-0.305(10)$ & 1.870(6) & $-0.50(3)$ & $-0.34(6)$ \\
22 & 1.0993(3) & $-0.515(4)$ & $-0.365(10)$ & 2.6320(40) & $-0.467(15)$ & $-0.270(30)$ & 1.865(2) & $-0.48(1)$ & $-0.30(2)$ \\
24 & 1.0992(3) & $-0.514(3)$ & $-0.364(8)$ & 2.6330(50) & $-0.470(15)$ & $-0.280(40)$ & 1.870(3) & $-0.495(20)$ & $-0.34(4)$ \\
26 & 1.0990(3) & $-0.511(3)$ & $-0.355(8)$ & 2.6320(20) & $-0.468(10)$ & $-0.270(20)$ & 1.867(3) & $-0.485(20)$ & $-0.31(4)$\\
28 & 1.0989(2) & $-0.509(2)$ & $-0.349(8)$ & 2.6330(20) & $-0.469(7)$ & $-0.270(15)$ & 1.868(2) & $-0.490(10)$ & $-0.32(2)$ \\
30 & 1.0988(2) & $-0.507(2)$ & $-0.341(8)$ & 2.6326(15) & $-0.469(5)$ & $-0.270(10)$ & 1.867(1) & $-0.483(7)$ & $-0.31(2)$ \\
\hline
Exact & 1.098612... &  $-0.49906$... &  $-0.30675$... &  2.633915... &   $-0.47810$... &  $-0.30477$...&  1.86626... &  $-0.48006$... & $-0.29809$...\\
\end{tabular}
\caption{Singularity parameters $\beta_c$, $A$, and $B$ for the triangular, honeycomb, and kagome lattices ferromagnetic Ising model obtained from different orders of the $c_v$-HTSE, compared with the exact solutions (see table \ref{tab0}).}
\label{tab2}
\end{table*}

We now turn to the triangular, honeycomb, and kagome lattices, keeping the ferromagnetic Ising model. They all present a finite-temperature phase transition with different non-universal parameters (see table \ref{tab0}). Our method works very well for the triangular lattice, with quality values close to 1 for all orders\cite{Supplemental}. On the other hand, for the honeycomb and kagome lattices the maximum of $Q$ oscillates between 0.5 and 0.7 depending on the HTSE order and the lattice\cite{Supplemental}. Moreover, these two lattices present wider peaks of $Q(\beta_c, A)$ which lead to a loss of precision in the evaluation of the parameters $\beta_c$, $A$, and $B$ in the end. Nevertheless, it is worth mentioning that having $50\%$ of coinciding PAs up to $\beta_c$ is a very good result since the PAs of the raw HTSE are all different at $\beta_c$.

The quality pictures (see figure \ref{fig1} and those in the supplemental\cite{Supplemental}) are qualitatively different from one lattice to another. It is not easy to determine the conditions under which some models possess large and sharp quality peaks or not. The hypothesis that a low value of $\beta_c$ (that decreases with the coordination number $p$ of the lattice roughly as $T_c \propto p-1$) makes the extrapolation method work better is not verified for the square and kagome lattices (both with a coordination of 4). Although they have about the same $\beta_c$, the first one has a sharp quality peak, and the second a wide one. On the other hand, bipartite (square and honeycomb) or non-bipartite (triangular and kaome) lattices can both either have sharp or wide quality peaks. Note that the bipartite property implies that odd terms of the HTSE are zero. Finally, we remark that the present method works better for the two Bravais lattices (square and triangular).

We summarize our results for these three lattices in table \ref{tab2}. Among the three parameters $\beta_c$, $A$ and $B$, it is always $\beta_c$ that has the smallest relative difference with the exact result. The $A$ and $B$ values for the triangular lattices converge to the exact values\cite{Supplemental}. On the other hand, for the honeycomb and kagome cases, the large uncertainties make it impossible to extrapolate to large $n$ (at least from the results of orders $n\leq 30$). Nevertheless, the reconstructed $c_v^r$ from any of these results is always in good agreement with the exact $c_v$\cite{Supplemental} (except very close to the critical point, because the divergence is not exactly at the same temperature).

Another interesting feature is the fact that, although the values of $\beta_c$ depend on the lattice, the values of the parameters $A$ and $B$ are always around the same values,  $-0.49$ for $A$ and and $-0.3$ for $B$ (see exact results in tables \ref{tab1} and \ref{tab2}). It would be indeed interesting to check if this is still the case for other more complicated lattices, or for other Ising transitions in general. 

\subsection{XXZ model}

We examine here the evolution of the Ising singularity when the model is interpolated between the Ising and Heisenberg model, i.e. when the parameter $\Delta$ of Eq.~\eqref{eq01} varies from 0 (Ising) to 1 (Heisenberg). In the Heisenberg limit, no discrete symmetry is broken in the ground state and the Mermin-Wagner theorem forbids the existence of finite-temperature phase transitions, but for any other value of $\Delta$ in the interval $[0,1)$ there is indeed a finite-temperature phase transition associated with the Ising order parameter. The value of the critical temperature $T_c$ is expected to decrease as $\Delta$ increases, and it has been suggested in ref. \cite{Ding92} that it behaves like the inverse of a logarithm near the Heisenberg point, $1/\ln|1-\Delta|$.

Since there are no exact results for this model, the HTSE coefficients have been calculated up to order 19 for the square lattice, 17 for the triangular case, 20 for the honeycomb case, and 18 for the kagome lattice. For the square and honeycomb lattices, we consider both ferro and antiferromagnetic interactions (since there is no frustration). In fact, in these cases where the lattice is bipartite, the transformation $\Delta \to -\Delta$ has no effect on the $c_v$-HTSE (this kind of transformation can be understood as a rotation of $\pi$ about the $z$-axis of one of the two sublattices). This implies that only the sign of the $z$ part is important in the evolution of the transition when $|\Delta|$ increases. On the other hand, for the triangular and kagome lattices only the ferromagnetic case has a finite-temperature phase transition. 

\begin{figure}[!t]
\begin{center}
\includegraphics*[width=0.45\textwidth]{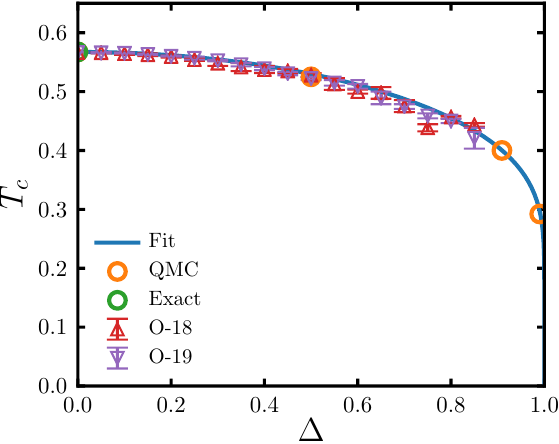}
\caption{Critical temperature $T_c$ as a function of $\Delta$ obtained with our method for the XXZ antiferromagnetic model (Eq.~\eqref{eq01}) on the square lattice, at HTSE orders 18 and 19. QMC results are taken from ref. \cite{Ding92}. The blue line is the fit of Eq.~\eqref{eqMod}. The green circle is the exact result in the Ising limit.}
\label{fig5}
\end{center}
\end{figure}

Figure \ref{fig5} shows our results for the square lattice in the antiferromagnetic case (the ferromagnetic case will be discussed below), using HTSE orders 18 (upper red triangles) and 19 (lower purple triangles). Near the Ising limit ($\Delta =0$), the uncertainties are always smaller than the symbol size, whereas they keep increasing when $\Delta$ increases, and the results starts to depend on the order. In fact, as $\Delta$ increases, the maximum value of the quality $Q(\beta_c, A)$ decreases while the domain of higher values widens, so it becomes difficult to determine precise values. The value of $Q$ decreases from 0.9 to 0.6 as $\Delta$ increases from 0 to 0.5, where our results are in good agreement with the QMC computations\cite{Ding92}. Indeed, at order 19 we find $T_c=0.522(3)$ whereas the QMC value is $T_c= 0.525(5)$\cite{Ding92}. The quality then continues to deteriorate to about 0.3 at $\Delta = 0.8$ and it becomes difficult to separate peaks from oscillations. However, our results seem to approach the next QMC point (see figure \ref{fig5}).

Using the exact limit in the Ising point as well as a vanishing derivative, and imposing the logarithmic behavior around $\Delta =1$, we propose a simple two-parameter function having the correct limits both for $\Delta =0$ and 1:
\begin{equation}
	T_c = T_c^{\rm Ising}\frac{1+a\Delta-b\Delta^2}{1-a\ln(1-\Delta)}
\label{eqMod}
\end{equation}
Fitting these parameters using the three QMC points\cite{Ding92} and the exact Ising value leads to $a=0.26$ and $b=0.11$. This function is is shown in figure \ref{fig5}, where we can see that it agrees with our current results.

\begin{figure}[!t]
\begin{center}
\includegraphics*[width=0.45\textwidth]{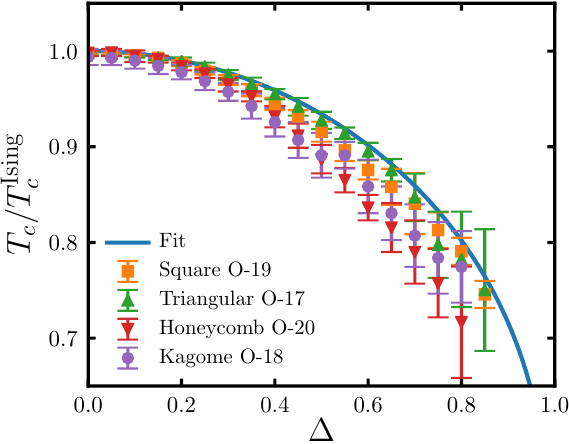}
\caption{Critical temperature $T_c$, normalized by the corresponding exact value in the Ising limit, as a function of $\Delta$ obtained with the present method for the ferromagnetic square, triangular, honeycomb, and kagome lattices. In blue the fit of Eq.~\eqref{eqMod} for the square antiferromagnetic case.}
\label{fig6}
\end{center}
\end{figure}

In figure \ref{fig6} we show the critical temperature, normalized by the exact critical temperature of the corresponding Ising case, for the square, triangular, honeycomb, and kagome lattices; using ferromagnetic interactions. Although the relative critical temperatures are very similar for all four lattices at low $\Delta$, larger differences appear above $\Delta = 0.2$. The honeycomb lattice, which is the one with the lowest coordination number, is the case where the relative critical temperature is the lowest, followed by the kagome and the square lattices and, finally, the triangular lattice which has the highest coordination number.

\begin{figure}[!t]
\begin{center}
\includegraphics*[width=0.45\textwidth]{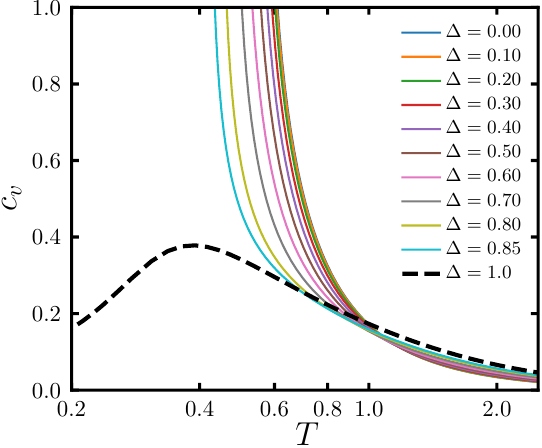}
\caption{Reconstructed specific heat $c_v$ as a function of $T$ for the square lattice ferromagnetic case. Several values of $\Delta$ are shown, from the Ising limit at $\Delta = 0.0$ to $\Delta=0.85$. For reference, we show in dashed line  the $c_v$ for the Heisenberg case obtained from the entropy method\cite{Bernu01}.}
\label{fig7}
\end{center}
\end{figure}

As was the case in the Ising models, the values of $A$ have larger uncertainties than those of $\beta_c$ and remain around the corresponding Ising values\cite{Supplemental}. As for the square lattice (see figure \ref{fig7}), uncertainties in $A$ have little effect in the reconstructed $c_v^r$ (Eq.~\eqref{eqrecon}), and again, variations in $\beta_c$ with $\Delta$ lead to the main effect on variations in $c_v$. The dashed black line stands for the Heisenberg specific heat ($\Delta = 1$), as obtained by the entropy method imposing a low-$T$  $c_v(T)\propto T^\alpha$ with $\alpha=1$. We can conclude from the figure \ref{fig7} that there must be an abrupt change in $c_v$ very close to $\Delta = 1$.

\section{Conclusion}
\label{secCon}

We have developed and tested an interpolation method using high-temperature series expansions that can be applied for finite-temperature phase transition systems with a logarithmic divergent specific heat $c_v$, as is the case for the two-dimensional Ising universality class. From the $c_v$-HTSE, we extract its logarithmic singular part to obtain the HTSE for the regular part of the $c_v$. These series depend on two parameters: the inverse of the critical temperature $\beta_c$ and the multiplying constant $A$ in front of the logarithmic term (see Eq.~\eqref{eqsing}). These parameters are determined by searching for the largest number of coinciding Pad\'e approximants. This method allows us to  accurately capture the critical behavior of the Ising model for several two-dimensional lattices, obtaining precise values of $\beta_c$ and good values for $A$ and $B$ (see tables \ref{tab1} and \ref{tab2}). With these values, we are able to reconstruct the $c_v$ above $T_c$ and we find a very good quantitative agreement with the exact solution already for orders as small as 14 (see figure \ref{fig3}). 

We then  turn to the XXZ Heisenberg model, where there are no exact results. We find that our method works in a wide range of anisotropies from $\Delta =0$ to 0.8. For example, at $\Delta = 0.5$ we find a precise value of the critical temperature $T_c$ which is in agreement with quantum Monte Carlo calculations. Our method then fails near the Heisenberg point where the critical temperature vanishes. The failure of the method is explicitly due to the fact that we do not find a set of parameters $\{\beta_c$,$A\}$ for which there are enough coinciding Pad\'e approximants; or that we do find a large domain of values for which there are some, so that $\beta_c$ and $A$ cannot be determined with precision. This is probably due to the complications that arise when there is a mixture of the singular behavior and the finite peak from the Heisenberg limit. Indeed, if the singular peak is at temperatures lower than the Heisenberg peak, removing the singular behavior would leave us with a series that must describe a Heisenberg-like $c_v$, and generally the Pad\'e approximants do not coincide for temperatures below $J/2$. A more advanced method would have to take  both peaks into account.

This limitation of our method is probably the reason why we did not get any convincing result for the antiferromagnetic $J_1-J_2$ model on the square lattice, where an Ising transition should appear for $J_2 > 0.6 J_1$ in the $S=\frac12$ quantum case. In this case, the transition temperature is small and it only increases for large values of $J_2$. But when $J_2$ is large, the transition is very subtle and difficult to capture\cite{Weber03, Capriotti04}.

At this stage, our method can be used to obtain important parameters related to phase transitions such as $\beta_c$, $A$, and $B$ in the case of logarithmic divergences. But it is possible to extend the method to other types of singularities, and other thermodynamic functions that show a singular behavior. Work in this direction is in progress, in particular for the case of three-dimensional Heisenberg models where $c_v$ shows a non-divergent cusp behavior (i.e., it reaches a maximum value with an infinite slope at $\beta_c$). In this case, our method could be used to estimate the critical exponent $\alpha$, which can not be calculated by the ratio or Dlog Pad\'e methods, and is usually only estimated indirectly by the relations between critical exponents\cite{Kuzmin19} or assumed to be the same as that predicted by field theory calculations\cite{Oitmaa96}. 

\section*{Acknowledgments}

The authors would like to thank Jesper Jacobsen for the fruitful discussions and for pointing us towards some relevant references regarding the Ising model, and Michael Kuz'min and Johannes Richter for fruitful discussions regarding the series expansions. This work was supported by the French Agence Nationale de la Recherche under Grant No. ANR-18-CE30-0022-04 LINK and by an \textit{Emergence(s)} project funded by the Paris city. 

\appendix

\section{Ising exact formulae}

\subsection{General formula}
\label{app1}
The free energy, $f$, of the Ising model on some two-dimensional lattices such as the square, triangular, honeycomb, and kagome lattices can be written as\cite{Plechko98}
\begin{align}
\label{EQ-f-GEN}
	\beta f&= f_0-\frac1{2n_s}\int_0^{2\pi}\!\!\!\!\int_0^{2\pi}\!\!\!\!\!\!dp\,dq\,
		\frac {\ln  \left( u-\,S_\delta(p,q) \right)}{4\pi^2}\\
	S_\delta(p,q)&=\frac{1}{2+\delta}\left(\cos p + \cos q + \delta\cos(p+q)\right)
\end{align}
where $f_0$ and $u$ are functions of $\beta$, and $n_s$ is the number of sites per unit cell, all depending on the lattice. We have $\delta=0$ for the square lattice and 1 for the triangular, honeycomb, and kagome lattices. The maximum value of $S_\delta(p,q)$ is 1 and the singular behavior occurs when the argument of the logarithm vanishes, that is when $S_\delta(p,q)=u=1$. In the following, we assume $u>1$.

The specific heat can then be calculated as
\begin{align}
	c_v&=-\beta^2\frac{d^2\beta f}{d\beta^2} \nonumber\\
		&=-\beta^2\frac{d^2f_0}{d\beta^2}+\frac {\beta^2}{8\pi^2n_s}\!\int_0^{2\pi}\!\!\int_0^{2\pi}\!\!dp\,dq\,X_\delta(p,q)
\end{align}
where
\begin{align}
\label{EQ-def-X}
		X_\delta(p,q)&=\frac{d}{d\beta}\frac{u'}{u-\,S_\delta(p,q)}\nonumber \\
		&=\frac{u''}{u-S_\delta(p,q)}-\frac{u'^2}{(u-S_\delta(p,q))^2}
\end{align}
where the prime indicates a derivative with respect to $\beta$. 
We have to calculate the following integrals 
\begin{align}
	I_k&=\int_0^{2\pi}\!\!\!\!\int_0^{2\pi}\!\!\frac {dp\,dq}{(u-S_\delta(p,q))^k}
\end{align}
for $k=$1 and 2, so that
\begin{align}
\label{EQ-def-C2}
	C_2&=\int_0^{2\pi}\!\!\!\!\int_0^{2\pi}dp\,dq\,X_\delta(p,q)
	=u''\,I_1-(u')^2\,I_2
\end{align}
and finally
\begin{align}
\label{EQ-Cv-singularity}
	c_v&=-\beta^2\frac{d^2f_0}{d\beta^2}+\frac {\beta^2}{8\pi^2n_s}\left(u''\,I_1-(u')^2\,I_2\right)
\end{align}

\subsubsection{The $I_1$ integral}
For $u>1$, we can perform the integration over $q$:
\begin{align}
	\int_0^{2\pi}\!\!\!\frac {dq}{u-S_\delta(p,q)}&=\frac{2\pi n_c}{\sqrt {(u_+- \cos p)(\um-\cos p)}}
\end{align}
where $u_\pm=2u+\delta(u+1)\pm \sqrt{1+\delta(6u+2)}$, with $\um>1$ and $\up>\um+2\sqrt{1+8\delta}$. Then the  integral over $p$ gives
\begin{align}
\label{EQ-case-k1}
	I_1&=\frac{8\pi(2+\delta)\K_0}{\sqrt {\up-1}\sqrt {\um+1}}
\end{align}
where ${\mathcal K}_0={\rm EllipticK}(z_0)$ is the elliptic function $K$ (Maple definition), and the argument is
\begin{align}
	z_0&=\frac{\sqrt2\sqrt{\up-\um}}{\sqrt{\up-1}\sqrt{\um+1}}
\end{align}
For $\delta=0$, we find $z_0=1/u$ and $I_1 = 8\pi\K_0/u$.

\subsubsection{The $I_2$ integral}
For $u>1$, the integrations over $q$ and $p$ give
\begin{align}
\nonumber
	\int_0^{2\pi}\!\!\!&\frac {dq}{(u-S_\delta(p,q))^2}=\frac {2\pi(2+\delta)^2(\cos p-(2+\delta)u)}{ (\up\!- \cos p)^{3/2}(\um\!-\!\cos p)^{3/2}}\\
\label{EQ-I2}
	I_2&=\int_0^{2\pi}\!\!\!\!\int_0^{2\pi}\!\!\!\!\frac {dp\,dq}{(u-S_\delta(p,q))^2}=C_\K K_0 + C_\E \E_0
\end{align}
where $\E_0 = {\rm EllipticE}(z_0)$ is the elliptic function $E$, and
\begin{align}
	C_\K&=4\delta C(\um-1)(\up+1)\\\
	C_\E&=C\left[(\up\!-\!\um)^2(\up\!+\!\um)\!-\!2\delta(\up^2\!+\!\um^2\!-\!2)\right]\\
	C&=\frac{4\pi\,(2+\delta)^2}{(\up-\um)^2(\um-1)(\up+1)\sqrt{\um+1}\sqrt{\up-1}}
\end{align}
For $\delta =0$, $C_\K=0$ and $C_\E=8\pi/(u^2-1)$. With these two integrals, $I_1$ and $I_2$, and the expressions for $f_0$, $u$, and $n_s$ for each lattice, the complete $c_v$ can be obtained at all temperatures from equation \ref{EQ-Cv-singularity}.

\subsubsection{Singularities}
The singular behavior arises when $u$ goes to 1, that is when $\um$ goes to 1 and $\up$ goes to 3 or 7 for $\delta=0$ or 1, respectively. In the limit $u\to1$, we define the small variable $\epsilon=u-1$. Then, to the first order in $\epsilon$, we have 
\begin{align}
	I_1&=\frac{2\pi (2+\delta)}{\sqrt{1+2\delta}}\left(-\ln \frac {a\epsilon}8+\frac\epsilon2\left(\ln\frac {a\epsilon}8+b\right)\right)\\
\label{EQ-def-a-0}
	a&=1,\quad\ b=1\quad \, (\delta=0)\\
\label{EQ-def-a-1}
	a&=\frac23,\quad b=\frac56\quad (\delta=1)
\end{align}
Whereas for the second integral we have 
\begin{align}
	I_2&=\frac{4\pi}\epsilon-4\pi\left(1+\frac12\ln\frac\epsilon{8}\right)+...\ (\delta=0)\\
	I_2&=\frac{2\pi\sqrt3}\epsilon-\frac{\pi\sqrt3}6 \left(6\ln\frac\epsilon{12}+11\right)+...\ (\delta=1)
\end{align}

Around the critical point $u$ is proportional to $(t-t_c)^2$, thus also to $(\beta_c-\beta)^2$. We set then, using the fact that $u>1$,
\begin{align}
	u(\beta)=1+\alpha_L(\beta_c-\beta)^2,
\end{align}
where $\alpha_L$ depends on the lattice. Then $\epsilon=u-1=\alpha_L(\beta_c-\beta)^2$ and $u'(\beta)=-2\alpha_L(\beta_c-\beta)=-2\sqrt{\alpha_L\epsilon}$ and $u''(\beta)=2\alpha_L$. Finally, equation \ref{EQ-def-C2} becomes
\begin{align}
\label{EQ-C2}
	C_2=-\frac{4\pi \alpha_L(2+\delta)}{\sqrt{1+2\delta}}\left(\ln\frac{a\alpha_L(\beta_c-\beta)^2}{8}+2\right)
\end{align}
Using the leading term, we find that the singularity in $c_v$ from equation \ref{EQ-Cv-singularity} behaves as
\begin{align}
\label{EQ-Cvsing}
	c_v\sim- \frac{\alpha_L\beta_c^2(2+\delta)}{2\pi n_s\sqrt{1+2\delta}}\ln(\beta_c-\beta)
\end{align}

\subsection{Square lattice}
For the square lattice we have $\delta=0$ and $n_s=1$, then the functions in equation \ref{EQ-f-GEN} are
\begin{align}
	f_0&=-\ln2+\ln(1-t^2)-\frac1{2n_s}\ln(2B)\\
	A&=(1+t^2 )^2\\
	B&=2t(t^2 - 1)\\
	u&=\frac A{2B}=\frac{(1+t^2)^2}{4t(1-t^2)}
\end{align}
where $t=\tanh\frac\beta4$. The critical point is for  $u=1$, that is $t_c=\sqrt2-1$ and $\beta_c=\ln(3+2\sqrt2)$. 
Around $\beta_c$ we have $\epsilon=u-1=\frac14(\beta_c-\beta)^2+...$, thus $\alpha_L=\frac14$ in Eq. \ref{EQ-C2}.
Around $t_c$ $B$ is constant and the contributions of $f_0$ to $c_v$ gives at $\beta_c$: $-\frac{\beta_c^2}{8}$, and finally, we find
\begin{align}
	c_v&=\frac{\beta_c^2}{2\pi}\left(-\ln(\beta_c-\beta)+\frac52\ln2 - 1- \frac\pi{4}\right)+ o(1)
\end{align}

\subsection{Triangular lattice}
For the triangular lattice we have $\delta = 1$ and $n_s=1$, then the functions in equation \ref{EQ-f-GEN} are
\begin{align}
	f_0&=-\ln2+\frac32\ln(1-t^2)-\frac1{2n_s}\ln(3B)\\
	A&=(t + 1)^2(t^4 - 2t^3 + 6t^2 - 2t + 1)\\
	B&=2t(t + 1)^2(t - 1)^2\\
	u&=\frac A{3B}=\frac16\left(t+\frac1t\right)+\frac23\left(\frac1{t-1}+\frac1{(t-1)^2}\right)
\end{align}
where $t=\tanh\frac\beta4$. The critical point is for  $u=1$, that is $t_c=2-\sqrt3$ and $\beta_c=\ln3$. Around $\beta_c$ we have $\epsilon=u-1=\frac34(\beta_c-\beta)^2+...$, thus $\alpha_L=\frac34$ in equation \ref{EQ-C2}. Around $t_c$, $B$ is constant and the contributions of $f_0$ to $c_v$ give at $\beta_c$: $-\frac38 \beta_c^2$, and finally we find
\begin{align}
	c_v&=\frac{3\sqrt3\beta_c^2}{4\pi}\left(-\ln(\beta_c-\beta)+2\ln2 - 1- \frac\pi{2\sqrt3}\right) + o(1)
\end{align}

\subsection{Honeycomb lattice}
For the honeycomb lattice we have $\delta =1$ and $n_s=2$, then the functions in equation \ref{EQ-f-GEN} are
\begin{align}
	f_0&=-\ln2+\frac34\ln(1-t^2)-\frac1{2n_s}\ln(3B)\\
	A&=1+3t^4\\
	B&=2t^2(1-t^2)\\
	u&=\frac A{3B}=\frac16\frac{1+3t^4}{t^2(1-t^2)}
\end{align}
where $t=\tanh\frac\beta4$. The critical point is for  $u=1$, that is $t_c=1/\sqrt3$, and $\beta_c=2\ln(2+\sqrt3)$. Around $\beta_c$ we have $\epsilon=u-1=\frac14(\beta_c-\beta)^2+...$, thus $\alpha_L=\frac14$ in equation \ref{EQ-C2}. Around $t_c$, $B$ is constant and the contributions of $f_0$ to $c_v$ gives at $\beta_c$: $-\frac{\beta_c^2}{24}$, and finally we find
\begin{align}
	c_v&=\frac{\beta_c^2\sqrt3}{8\pi}\left(-\ln(\beta_c-\beta) +\frac{\ln48}2 - 1 - \frac{\pi\sqrt3}{9}\right) + o(1)
\end{align}

\subsection{Kagome lattice}
For the kagome lattice we have $\delta=1$ and $n_s=3$, then the functions in equation \ref{EQ-f-GEN} are
\begin{align}
	f_0&=-\ln2-\frac1{2n_s}\ln(3B)\\
	A&=\frac{(z-1)^4+3}{(t - 1)^6(t + 1)^2}t^4,\qquad z=t+\frac1t\\
	B&=\frac{2(t^2 + 1)t^2}{(t - 1)^4(t + 1)^2}\\
	u&=\frac A{3B}=1+\frac{(z^2-2z-2)^2}{6t^2(1+t^2)(1-t)^2}
\end{align}
where $t=\tanh\frac\beta4$. The critical point is for  $u=1$, that is $t_c=\frac12(\sqrt3+1-\sqrt23^{1/4})$, and $\beta_c=\ln(3+2\sqrt3)$. Around $\beta_c$ we have $\epsilon=u-1=\frac34(\beta_c-\beta)^2+...$, thus $\alpha_L=\frac34$ in equation \ref{EQ-C2}. Around $t_c$ $B$ is constant and the contributions of $f_0$ to $c_v$ gives at $\beta_c$: $-\frac{\beta_c^2(3-\sqrt3)}{24}$, and finally we find
\begin{align}
	c_v&=\frac{\beta_c^2\sqrt3}{4\pi}\left(-\ln(\beta_c\!-\!\beta) \!+\! 2\ln2 \!-\! 1 \!-\! \frac{\pi(\sqrt3-1)}{6}\right)\!\! +\!o(1)
\end{align}

\bibliography{papers}

\end{document}